\documentclass[reprint]{revtex4-1} 
\usepackage{graphicx}  
\usepackage{amsfonts}
\usepackage{geometry}
\usepackage{tabulary}
\usepackage{color}
\usepackage[normalem]{ulem}

\begin{document}
\title{Discovering the Building Blocks of Atomic Systems using Machine Learning}
\author{Conrad W. Rosenbrock}
\affiliation{Department of Physics and Astronomy, Brigham Young
  University, Provo, Utah 84602, USA.}
\author{Eric R. Homer}
\affiliation{Department of Mechanical Engineering, Brigham Young
  University, Provo, Utah 84602, USA.}
\author{G\'abor Cs\'anyi}
\affiliation{Engineering Laboratory, University of Cambridge, Trumpington Street, Cambridge CB2 1PZ, United Kingdom}
\author{Gus L. W. Hart}
\affiliation{Department of Physics and Astronomy, Brigham Young
  University, Provo, Utah 84602, USA.}

\date{\today}

\begin{abstract}
  Machine learning has proven to be a valuable tool to approximate functions in high-dimensional spaces. Unfortunately, analysis of these models to extract the relevant physics is never as easy as applying machine learning to a large dataset in the first place. Here we present a description of atomic systems that generates machine learning representations with a direct path to physical interpretation. As an example, we demonstrate its usefulness as a universal descriptor of grain boundary systems. Grain boundaries in crystalline materials are a quintessential example of a complex, high-dimensional system with broad impact on many physical properties including strength, ductility, corrosion resistance, crack resistance, and conductivity. In addition to modeling such properties, the method also provides insight into the physical "building blocks" that influence them. This opens the way to discover the underlying physics behind behaviors by understanding which building blocks map to  particular properties. Once the structures are understood, they can then be optimized for desirable behaviors.
\end{abstract}

\maketitle

As scientists continue to press for a deeper understanding on the natural world, they are eventually confronted with the sheer enormity of their task. While interactions between small, isolated components can be studied experimentally and then modeled, real-world systems include exponentially more complexity, and approximate, statistical methods are necessary in the quest for deeper understanding. Machine learning is a powerful statistical tool for extracting correlations from high-dimensional datasets; unfortunately, it often suffers from a lack of interpretability. Researchers can create models that approximate the physics well enough, but the physical intuition usually provided by models may be hidden within the complexity of the model (the black-box problem). Here we present a general method for representing atomic systems for machine learning so that there is a clear path to physical interpretation or the discovery of those ``building blocks'' that govern the properties of these systems.

We choose to demonstrate the method for crystalline interfaces because of their inherent complexity, high-dimensionality, and broad impact on many physical properties. Crystalline building blocks are well known and can be classified by a finite set of possible structures. 
Disordered atomic structures on the other hand are difficult to classify and there is no well-defined set of possible structures or building blocks. 
Furthermore, these disordered atomic structures often exhibit an oversized influence on material properties because they break the symmetry of the crystals.
Crystalline interfaces, more commonly called grain boundaries (GBs), are excellent examples of disordered atomic structures that exert significant influence on a variety of material properties including strength, ductility, corrosion
resistance, crack resistance, and conductivity \cite{Hall:1951cy,
  Petch:1953ws, Hansen:2004bn, Chiba:1994ur, Fang:2011ej,
  Shimada:2002jn, Lu:2004gh, Bagri:2011ip, Meyers:2006co}. They have macroscopic, 
crystallographic degrees of freedom that constrain the configuration between the two adjoining crystals \cite{Wolf:1992ve,Sutton:1995ux}. GBs also
have microscopic degrees of freedom that define the atomic structure of the GB \cite{Olmsted:2009ge, Cantwell:2013cu, Han:2015dhb, Dillon:2016uv}. While often classified experimentally using the crystallography, the crystallography is only a constraint, and it is the atomic structure that controls the GB properties. 

In this article, we examine the local atomic environments of GBs in an effort to discover their building blocks and influence on
material properties. This is achieved by machine learning on the space
of the atomic environments to make property predictions of GB energy,
temperature-dependent mobility trends, and shear coupling. The
implications of the work are significant; despite the immense number
of degrees of freedom, it appears that GBs in face-centered cubic Nickel are
constructed with a relatively small set of local atomic
environments. This means that the space of possible GB structures is
not only searchable, but that it is possible to find the atomic
environments that give desired properties and behaviors. We emphasize that in addition to being successful for modeling GBs, the methodology presented here could be applied generally to many atomic systems.

Atomic structures in GBs have been examined for decades using a variety of 
structural metrics \cite{Weins:1969dp, Ashby:1978ul, Gleiter:1982km, Frost:1982vc, Sutton:1989vz, Wolf:1990fk, Tschopp:2007wn, Tschopp:2007hr, Spearot:2008bq,  Olmsted:2009ge, Banadaki:2017dk} with the goal of obtaining structure-property relationships \cite{Read:1950um, Frank:1953fb, Bilby:1955jv, Wolf:1992ve,
  Sutton:1995ux, Wolf:1990ud, Wolf:1990fm, Yang:2010bs}. 
  Each of the efforts has provided unique insight, but none have given universal atomic structure-property relationships based on the large
number of possible atomic structures that GBs take, and their relationship with
specific material properties. 

Large databases of GB structures have produced property trends \cite{Olmsted:2009ge, Olmsted:2009in, Homer:2013ce, Homer:2014hr} and macroscopic \emph{crystallographic} structure-property relationships \cite{Bulatov:2014bz, Homer:2015ie}, but no \emph{atomic} structure-property relationships. 
Machine learning of GBs by Kiyohara \emph{et al}. \cite{Kiyohara:2015wb} has been used to make predictions of GB energy from atomic structures, but we are still left without an understanding of what is important in making the predictions, and how that affects our understanding of the underlying physics and the building blocks that control properties and behaviors.

\begin{figure*}[htp]
\centerline{\includegraphics[width=\textwidth]{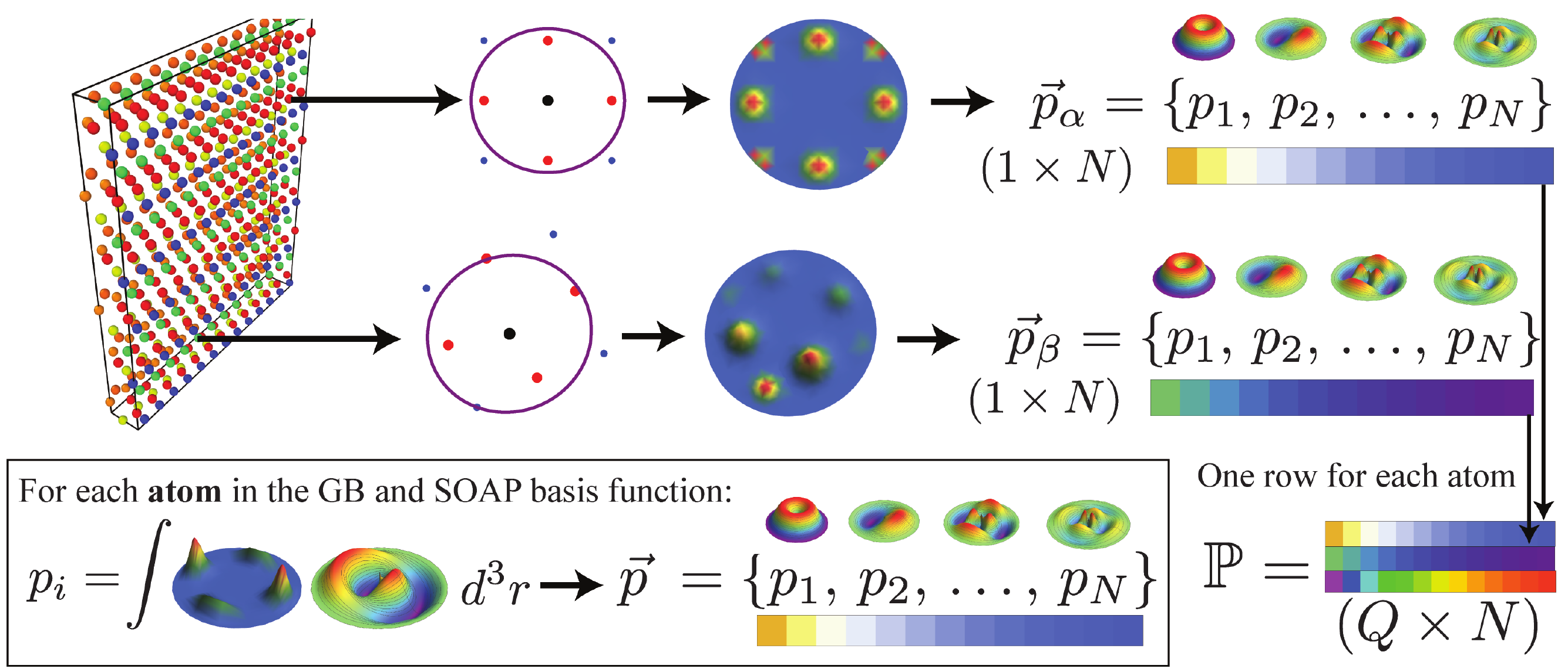}}
\caption[]{\label{fig:ConstructSOAPMatrix} Illustration of the process
  for extracting a SOAP matrix $\mathbb{P}$ for a single GB. Given a
  single atom in the GB, we place a Gaussian particle density function
  at the location of each atom within a local environment sphere
  around the atom. Next, the total density function produced by the
  neighbors is projected into a spectral basis consisting of radial
  basis functions and the spherical harmonics, as shown in the boxed region. Each basis function
  produces a single coefficient $p_i$ in the SOAP vector $\vec{p}$ for
  the atom, the magnitude of which is represented in the figure by the colors of the arrays. Once a SOAP vector is available for all $Q$ atoms in the
  GB, we collect them into a single matrix $\mathbb{P}$ that represents the
  GB. A value of $N=3250$ components in $\vec{p}$ is representative
  for the present work.  } 
\end{figure*}

To examine atomic structures, we adopt a descriptor for single-species grain
boundaries based on the Smooth Overlap of Atomic Positions (SOAP)
descriptor \cite{Bartok:2010fj, Bartok:2013cs}. The SOAP descriptor
uses a combination of radial and spherical spectral bases, including
spherical harmonics. It places Gaussian density distributions at the
location of each atom, and forms the spherical power spectrum
corresponding to the neighbor density. The descriptor can be expanded
to any accuracy desired and goes smoothly to zero at a finite distance
(compact support). 

The SOAP descriptor has the following qualities that make it ideal for Local Atomic Environment (LAE) characterization. Specifically, within GBs, the SOAP descriptor 1) is agnostic to the grains' specific underlying lattices (including the loss of periodicity at the GB); 2) has invariance to global translation,
global rotation, and permutations of identical atoms; 3) leads to a
metric that is smooth and differentiable. Assessing the similarity
between two local environments in the SOAP vector space requires only
a simple dot product. In GBs, the SOAP descriptor has advantages over other structural metrics in that it requires no predefined set of structures, and a small change in atomic positions won't lead to a drastic redefinition of the SOAP
environment \cite{Sutton:1989vz, Tschopp:2007hr, Spearot:2008bq, Ashby:1978ul, Gleiter:1982km}.

Figure \ref{fig:ConstructSOAPMatrix} illustrates the process for
determining the SOAP descriptor for a GB. First, GB atoms and some
surrounding bulk atoms are isolated from their surroundings; a SOAP
descriptor for each atom in the set is calculated and represented as a
vector of coefficients.  The matrix of these vectors, one for each
LAE, is the full SOAP representation for
each GB. The SOAP vector can be expanded to resolve any desired
features. For the present work, a cutoff distance of 5\AA $\,$ and vector of
length 3250 elements produced good results. The computed GBs studied
in this work are the 388 Ni GBs created by Olmsted, Foiles, and Holm
\cite{Olmsted:2009ge}, using the Foiles-Hoyt embedded atom method
(EAM) potential \cite{Foiles:2006cp}.

\begin{figure*}[htbp]
\centerline{\includegraphics[width=\textwidth]{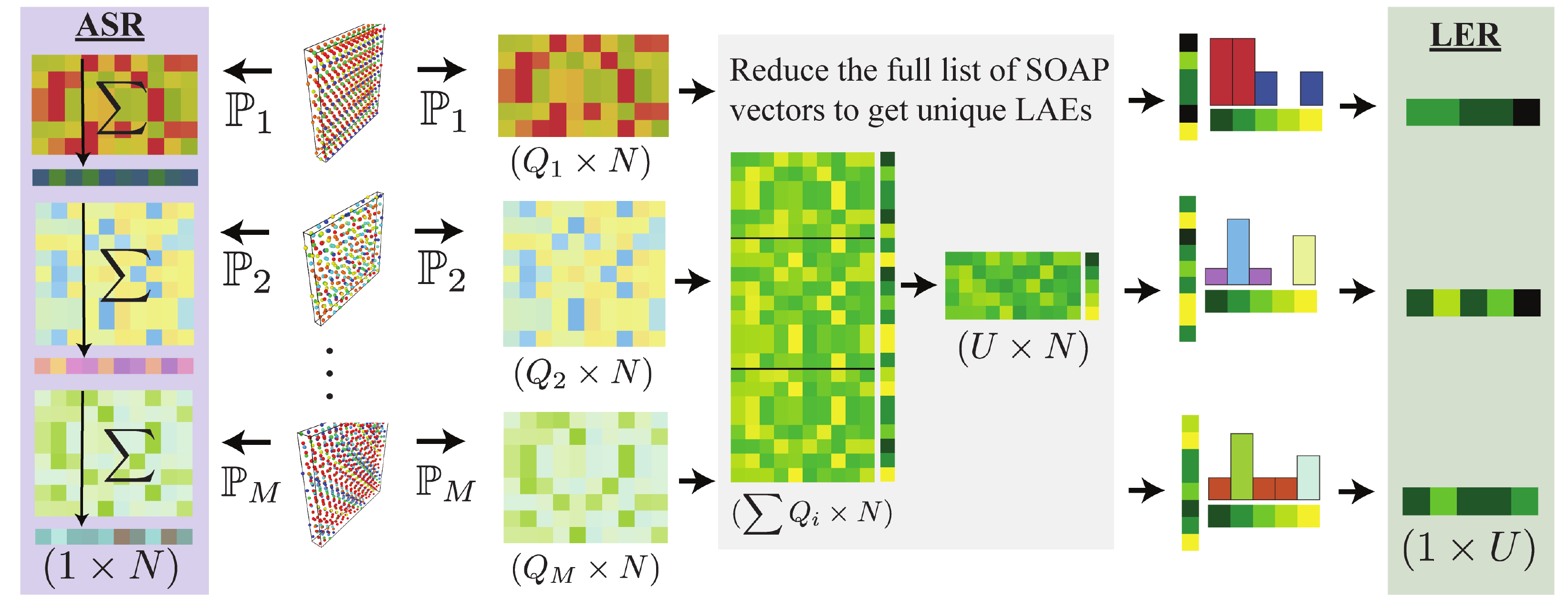}}
\caption[]{\label{fig:LERASR} Illustration of the process for
  construction of the ASR and LER for a collection of GBs. First, a
  SOAP matrix $\mathbb{P}$ is formed (as shown in Figure
  \ref{fig:ConstructSOAPMatrix}). ASR: A sum down each of the $Q$ columns in
  the matrix produces an averaged SOAP vector that is representative
  of the whole GB. The ASR feature matrix is then the collection of averaged SOAP vectors for all $M$ GBs of interest ($M \times N$).
  LER: The SOAP vectors from all $M$ GBs in the
  collection are grouped together and reduced to a set $U$ of
  \emph{unique} vectors using the SOAP similarity metric, of which each unique vector represents a unique LAE. A histogram
  can then be constructed for each GB counting how many examples of
  each unique vector are present in the GB. This histogram produces a
  new vector (the LER) of fractional abundances, whose components sum
  to 1. The LER feature matrix is then the collection of histograms of unique LEA  for the $M$ GBs in the collection ($M \times U$). }
\end{figure*} 

We investigate two approaches for applying machine learning to the GB
SOAP matrices. For the first option, we average the SOAP vectors, or
coefficients, of all the atoms in a single GB to obtain one averaged
SOAP vector that is a measure of the whole GB as shown in Figure
\ref{fig:LERASR}. In other words, it is a single description of the average LAE for the whole GB structure. We refer to this single averaged vector as the Averaged SOAP Representation (ASR). The ASR for a collection of GBs becomes the
feature matrix for machine learning. 

Alternatively, we can compile an exhaustive set of \emph{unique} LAEs
by comparing the environment of every atom in every GB to all other
environments using the SOAP similarity metric and a numerical
similarity parameter (see Figure \ref{fig:LERASR}). In the
present work, 800,000 LAEs from the atoms in 388 GBs are reduced to
145 unique LAEs. This is a considerable reduction in dimensionality
for a machine learning approach. More importantly, these 145 unique
LAEs mean that there may be a relatively small, finite set of LAEs
used to construct every possible GB in Ni. Using the reduced set of
unique LAEs, we represent each GB as a vector whose components are the
fraction of each globally unique LAE in that GB. This GB
representation is referred to as the Local Environment Representation
(LER), and the matrix of LER vectors representing a collection of GBs
is also a feature matrix for machine learning. \emph{The 145 unique
  LAEs give a bounded 145-dimensional space, which is a significant
  improvement over the 3*800,000-dimensional configurational space.}

These two approaches are used because they are complementary: physical
quantities such as energy, mobility, and shear coupling are best
learned from the ASR, while physical interpretability is accessible
using the LER, with only marginal loss in predictive power. Because we
desire to discover the underlying physics and not just provide a
black-box for property prediction, we use the LER to deepen our
understanding of which LAEs are most important in predicting material properties such as mobility and shear coupling. 

A summary of the machine learning predictions by the various methods is provided
in Table \ref{table:ASRLERPerformance}. Machine learning was performed using the
ASR and LER descriptions of the GBs and the properties of interest for the
learning and prediction are GB energy, temperature-dependent mobility, and shear
coupled GB migration (obtained from the computed Ni GBs). Table \ref{table:ASRLERPerformance} also includes the results of attempting to predict these properties by "educated" random guessing using knowledge of the statistical behavior of the training set. In all cases, the machine learning predictions are significantly better than random draws from distributions appropriately matched to the training data.

GB energy is measured as the excess energy of a grain boundary relative to the bulk energy as a result of the irregular structure of the atoms in the GB \cite{Olmsted:2009ge,Tadmor:2011tb}. GB energy is a static property of the system measured at 0\,K, and all atomistic structures examined in
the machine learning are the 0\,K structures associated with this calculation. 

Temperature-dependent mobility and shear coupled GB migration are
two dynamic properties related to the behavior of a GB when it migrates. The
temperature-dependent mobility trend classifies each GB as having (i)
\emph{thermally activated}, (ii) \emph{athermal}, (iii) \emph{thermally damped}
mobility depending on whether the mobility of the GB (related to the migration
rate) increases, is constant, or decreases with increasing temperature
\cite{Homer:2014hr}. GBs that do not move under any of these conditions are
classified as being (iv) \emph{immobile}. In addition, when GBs migrate, they
can also exhibit a coupled shear motion, in which the motion of a GB normal to
its surface couples with lateral motion of one of the two crystals
\cite{Cahn:2006gt,Homer:2013ce}. GBs are then classified as either exhibiting
shear coupling or not. 

GB energy is a continuous quantity, while temperature
dependent mobility trend and shear coupling are classification
properties. Additional details regarding these properties are available in the
publications pertaining to their measurements
\cite{Olmsted:2009ge,Homer:2014hr,Homer:2013ce}. For the mobility and shear
coupling classification, the dataset suffered from imbalanced classes; we used
standard machine learning resampling techniques to help mitigate the problem
\cite{Han:2005:BNO:2141202.2141297,Nguyen:2011:BOI:1972030.1972031,lemaitre2016imbalanced}.

\begin{figure}[htbp]
\centerline{\includegraphics[width=0.5\textwidth]{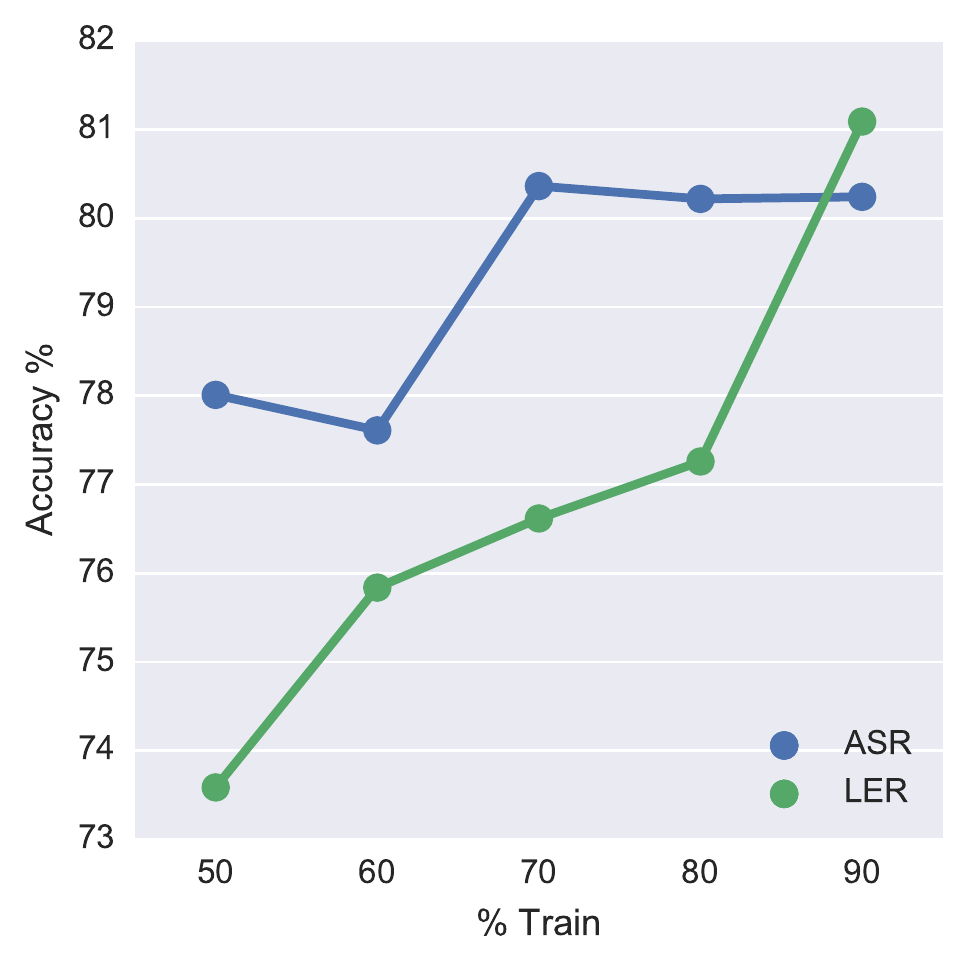}}
\caption[]{\label{fig:LearningRate} Learning rate of ASR vs. LER for
  mobility classification as a function of the specified split of training vs. validation
  data. The accuracy was calculated over 25
  independent fits. It appears that the LER accuracy increases faster with more
  data, though a larger dataset is necessary to confidently establish
  this point.}
\end{figure}

Unfortunately, the size of the dataset is a limiting factor in the
performance of the machine learning models. In Table
\ref{table:ASRLERPerformance}, we used only half of the available 388
GBs for training. As we increase the amount of training data given to
the machine, the learning rates change as shown in Figure
\ref{fig:LearningRate}. Although it is common practice  to
use up to 90\% of the available data in a small dataset for training (with suitable cross
validation), we chose to use a lower (pessimistic) split to guarantee
that we are not overfitting to non-physical features. Larger datasets
would certainly improve the models and our confidence in the physics
they illuminate.

\begin{table*}[b]
\caption[]{Predictive performance of the machine learning models trained
  on the ASR and LER representation respectively. The models were
  trained on 50\% (194) of the available 388 GBs and then validated on
  the remaining 194 GBs that the model had never seen. Percent
  error is relative to the mean. Error bars represent the standard
  deviation over 50 independent, random samplings (including different
  combinations of the 50\% split), and re-fits of the dataset. For the
  “Random” column, energies were guessed by drawing values from a normal
  distribution that had the same mean and standard deviation as the 50\%
  training data, and then compared to the actual energies in the
  validation data. For the classification problems, random choices from
  the 50\% training data class labels were compared to the validation
  data.
  } 
\begin{tabular*}{0.837\textwidth}{  lccc }
  \hline
  Property & ASR & LER & Random \\
  \hline
  GB Energy & 89.2 $\pm$ 0.7\% & 88.5 $\pm$ 0.9\% & 70.4 $\pm$ 1.6\%  \\ 
  Temperature Dependent Mobility Trend & 77.4 $\pm$ 2.5\% & 74.3 $\pm$ 2.7\% & 38.5 $\pm$ 2.0\% \\ 
  Shear Coupling & 61.3 $\pm$ 0.6\% & 61.4 $\pm$ 0\% & 52.0 $\pm$ 2.5\% \\
  \hline
\end{tabular*}
\label{table:ASRLERPerformance}
  
\end{table*}

For small datasets, ASR does slightly better in predicting energy and
temperature-dependent mobility trend; ASR and LER are essentially equivalent for
shear coupling. However, the ASR suffers from a lack of interpretability because
1) its vectors and similarity metric live in the abstract SOAP space, which is
large and less intuitive; 2) the results reported for ASR were obtained using
machine learning algorithms that are not easily interpretable \footnote{Details
  on the algorithm types and other details are included in the supplementary
  information.  }. The LER, on the other hand, \emph{has direct analogues in LAEs that can be analyzed in their original physical context}. The best-performing
algorithms for the LER are gradient-boosted decision trees, which lend
themselves to easy interpretation. Even at slightly
lower accuracy, the physical insights generated by the LER make it the superior
choice.

\begin{figure*}[htbp]
\centerline{\includegraphics[width=\textwidth]{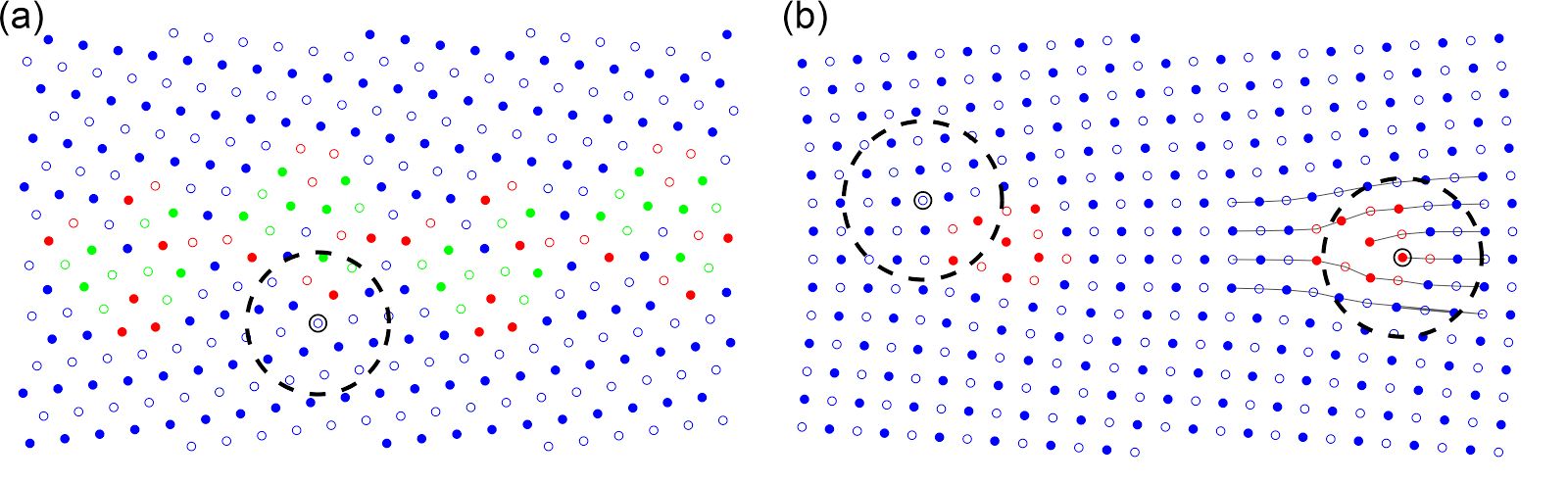}}
\caption[]{\label{fig:ImportantLAEs} Illustration of important LAEs
  for classifying thermally activated GB mobility, as identified in
  two different GBs. The GB shown in (a) is a $\Sigma$51a (16.1$^{\circ}$ symmetric
  tilt about the [110] axis, $\{1\: \bar{1}\: 10\}$ boundary planes)
  GB, and has one LAE identified.The LAE shown in (a) has a relative
  importance of 3\% over the entire system and includes a leading
  partial dislocation that originates from the GB. The GB shown in (b)
  is a $\Sigma$85a (8.8$^{\circ}$ symmetric tilt about the [100] axis,
  $\{0\: \bar{1}\: 13\}$ boundary planes) GB, and has two LAEs
  identified.  The leftmost LAE has a relative importance of 9\% (for
  all GBs in the dataset) but its structural importance is not
  immediately clear, offering an exciting opportunity to discover new
  physics. The second LAE in (b) encloses edge dislocations, which are
  regularly spaced to form a tilt GB, (relative importance of 2.7\%
  across all GBs). The open and filled circles denote atoms on the two
  unique stacking planes along the [100] or [110] direction. The atoms
  are colored according to common neighbor analysis (CNA) such that
  blue, green, and red atoms have a local environment that is FCC,
  HCP, or unclassifiable.  }
\end{figure*}

In Figure \ref{fig:ImportantLAEs}, we plot three of the top ten most
important environments for determining whether a grain boundary will exhibit
thermally activated mobility or not. These most important
LAEs are classified as such because their presence or absence
in any of the GBs in the \emph{entire dataset} is highly correlated
with the decision to classify them as thermally activated or
not. Since such global correlations must be true for all GBs in the
system, we assume that they are tied to underlying physical processes.

Figure \ref{fig:ImportantLAEs}a shows a LAE centered around a leading partial
dislocation. GBs with partial dislocations emerging from the structure
have been associated with thermally activated mobility and immobility,
depending upon their presence in simple or complex GB structures
\cite{Homer:2013ce}; in addition, these structures have also been
associated with shear coupled motion or the lack thereof. We now know
that there is a strong correlation between the presence of these LAEs
and their mobility type, though the presence of other structures is
also important in the determination of the exact mobility
type. \emph{This LAE was presented on equal footing with all others in
  the feature matrix that trained the machine. In the training, it was
  selected as important and we can easily see that it has relevant
  physical meaning}. 

In Figure \ref{fig:ImportantLAEs}b another LAE has obvious physical
meaning as it captures edge dislocations in the environment of the
selected atom. Interestingly, arrays of these edge dislocations, as in
Figure \ref{fig:ImportantLAEs}b, are the basis for the energetic
structure-property relationship of the Read-Shockley model
\cite{Read:1950um}.

Thus, in these first two cases, we see that the LER approach discovers
well-known, and physically important structures or defects that are
commonly identified in metallic structures. Perhaps even more
interesting is the second LAE in Figure \ref{fig:ImportantLAEs}b,
which has the highest relative importance of all ($\approx$9\%). This LAE
includes mostly perfectly structured FCC atoms, though it also
includes the edge of a defect. While this structure is not immediately
identified with any known metallic defect, we know that it is highly
correlated with thermally activated mobility across \emph{all} the GBs in the dataset. This offers an exciting avenue to discover new
mechanisms and structures governing these physical properties. The
physical nature of those LAEs that we already understand suggests that
these are the “building blocks” underlying important physical
properties and that we may be on the precipice of understanding the
atomic building blocks of GBs.

Despite the formidable dimensionality of a raw grain boundary system,
machine learning using SOAP-based representations makes the problem
tractable. In addition to learning useful physical properties, the
models provide access to a finite set of physical “building blocks”
that are correlated with those properties throughout the high-dimensional GB space. Thus, the machine learning is not just a black box
for predictions that we don't understand. The work shows that
analyzing big data regarding materials science problems can provide
insight into physical structures that are likely associated with
specific mechanisms, processes, and properties but which would
otherwise be difficult to identify.  Accessing these building blocks
opens a broad spectrum of possibilities. For example, the reduced
space can now be searched for extremal properties that are unique
(i.e., “special” grain boundaries). Poor behavior in certain
properties can be compensated for by searching for combinations of
other properties. In short, a path is now available to develop methods
that optimize grain boundaries (at least theoretically) at the
atomic-structure scale. This is the beginning of atomic
structure-property relationships that are applicable to all possible
GB structures. These methods may also provide a route to connect the
crystallographic and atomic structure spaces so that existing
expertise in the crystallographic space can be further optimized
atomistically or vice versa.

While this is exciting within grain boundary science, the methodology presented here has general applicability to any atomistic system with many degrees of freedom. The physical interpretability of the machine learning representations, in terms of atomic environments, will also transfer well to new applications. This can lead to increased physical intuition across many fields of research that are confronted with the same, formidable complexity as seen in grain boundary science.

\section*{Author Contributions}

CWR conceived the idea, performed all the calculations, and wrote a significant portion of the paper.
ERH was responsible for interpretation of the results and guidance of the project, and also wrote a significant portion of the paper. GC provided code, guidance and expertise in applying SOAP to the GBs.
GLWH contributed many ideas and critique to help guide the project, and helped write the paper.

\section*{Acknowledgments}

CWR and GLWH were supported under ONR (MURI N00014-13-1-0635). ERH is supported by the U.S. Department of Energy, Office of Science, Basic Energy Sciences under Award \#DE-SC0016441.

\section*{Additional Information}

Additional details about the machine learning models and data are described in the accompanying supplementary information. The feature matrices and code to generate them will be made available to individuals upon request. 

\section*{Competing Financial Interests}
The authors declare no competing financial interests.


\begin{thebibliography}{9}

\bibitem{Ashby:1978ul}
M~F Ashby, F~Spaepen, and S~Williams.
\newblock {Structure of Grain Boundaries Described as a Packing of Polyhedra}.
\newblock {\em Acta Metall Mater}, 26(11):1647--1663, 1978.

\bibitem{Bagri:2011ip}
Akbar Bagri, Sang-Pil Kim, Rodney~S Ruoff, and Vivek~B Shenoy.
\newblock {Thermal transport across Twin Grain Boundaries in Polycrystalline
  Graphene from Nonequilibrium Molecular Dynamics Simulations}.
\newblock {\em Nano Lett.}, 11(9):3917--3921, September 2011.

\bibitem{Banadaki:2017dk}
Arash~D Bandaki and Srikanth Patala.
\newblock A three-dimensional polyhedral unit model for grain boundary
  structure in fcc metals.
\newblock {\em npj Computational Materials}, 2017.

\bibitem{Bartok:2013cs}
Albert~P Bart{\'o}k, Risi Kondor, and G{\'a}bor Cs{\'a}nyi.
\newblock {On representing chemical environments}.
\newblock {\em Phys Rev B}, 87(18):184115, May 2013.

\bibitem{Bartok:2010fj}
Albert~P Bart{\'o}k, Mike~C Payne, Risi Kondor, and G{\'a}bor Cs{\'a}nyi.
\newblock {Gaussian Approximation Potentials: The Accuracy of Quantum
  Mechanics, without the Electrons}.
\newblock {\em Phys Rev Lett}, 104(13):136403, April 2010.

\bibitem{Bilby:1955jv}
B~A Bilby, R~Bullough, and E~Smith.
\newblock {Continuous Distributions of Dislocations: A New Application of the
  Methods of Non-Riemannian Geometry}.
\newblock {\em Proc Roy Soc A-Math Phy}, 231(1185):263--273, August 1955.

\bibitem{Bulatov:2014bz}
Vasily~V Bulatov, Bryan~W Reed, and Mukul Kumar.
\newblock {Grain boundary energy function for fcc metals}.
\newblock {\em Acta Materialia}, 65:161--175, 2014.

\bibitem{Cahn:2006gt}
John~W Cahn, Yuri Mishin, and Akira Suzuki.
\newblock {Coupling grain boundary motion to shear deformation}.
\newblock {\em Acta Materialia}, 54(19):4953--4975, 2006.

\bibitem{Cantwell:2013cu}
Patrick~R Cantwell, Ming Tang, Shen~J Dillon, Jian Luo, Gregory~S Rohrer, and
  Martin~P Harmer.
\newblock {Grain boundary complexions}.
\newblock {\em Acta Materialia}, September 2013.

\bibitem{Chiba:1994ur}
A~Chiba, S~Hanada, S~Watanabe, T~Abe, and Obana T.
\newblock {Relation Between Ductility and Grain-Boundary Character
  Distributions in Ni3al}.
\newblock {\em Acta metallurgica et Materialia}, 42(5):1733--1738, May 1994.

\bibitem{Dillon:2016uv}
Shen~J Dillon, Kaiping Tai, and Song Chen.
\newblock {The importance of grain boundary complexions in affecting physical
  properties of polycrystals}.
\newblock {\em Current Opinion in Solid State and Materials Science}, June
  2016.

\bibitem{Fang:2011ej}
T~H Fang, W~L Li, N~R Tao, and K~Lu.
\newblock {Revealing Extraordinary Intrinsic Tensile Plasticity in Gradient
  Nano-Grained Copper}.
\newblock {\em Science}, 331(6024):1587--1590, March 2011.

\bibitem{Foiles:2006cp}
Stephen~M Foiles and J~J Hoyt.
\newblock {Computation of grain boundary stiffness and mobility from boundary
  fluctuations}.
\newblock {\em Acta Materialia}, 54(12):3351--3357, 2006.

\bibitem{Frank:1953fb}
F~C Frank.
\newblock {Martensite}.
\newblock {\em Acta Metall Mater}, 1(1):15--21, January 1953.

\bibitem{Frost:1982vc}
H~J Frost, M~F Ashby, and F~Spaepen.
\newblock {A Catalogue of [100], [110], and [111] Symmetric Tilt Boundaries in
  Face-Centered Cubic Hard Sphere Crystals}.
\newblock {\em Harvard Division of Applied Sciences}, pages 1--216, June 1982.

\bibitem{Gleiter:1982km}
H~Gleiter.
\newblock {On the structure of grain boundaries in metals}.
\newblock {\em Mater Sci Eng}, 52(2):91--131, February 1982.

\bibitem{Hall:1951cy}
E~O Hall.
\newblock {The Deformation and Ageing of Mild Steel: III Discussion of
  Results}.
\newblock {\em Proc. Phys. Soc. B}, 64(9):747--753, 1951.

\bibitem{Han:2005:BNO:2141202.2141297}
Hui Han, Wen-Yuan Wang, and Bing-Huan Mao.
\newblock Borderline-smote: A new over-sampling method in imbalanced data sets
  learning.
\newblock In {\em Proceedings of the 2005 International Conference on Advances
  in Intelligent Computing - Volume Part I}, ICIC'05, pages 878--887, Berlin,
  Heidelberg, 2005. Springer-Verlag.
  
\bibitem{Han:2015dhb}
J Han, V Vitek, and D~J Srolovitz
\newblock {The interplay between grain boundary structure and defect sink/annealing
  behavior}.
\newblock {\em IOP Conference Series: Materials Science and Engineering},
  89(1):012004, 2015.

\bibitem{Hansen:2004bn}
Niels Hansen.
\newblock {Hall{\textendash}Petch relation and boundary strengthening}.
\newblock {\em Scripta Mater}, 51(8):801--806, October 2004.

\bibitem{Homer:2013ce}
Eric~R Homer, Stephen~M Foiles, Elizabeth~A Holm, and David~L Olmsted.
\newblock {Phenomenology of shear-coupled grain boundary motion in symmetric
  tilt and general grain boundaries}.
\newblock {\em Acta Materialia}, 61(4):1048--1060, February 2013.

\bibitem{Homer:2014hr}
Eric~R Homer, Elizabeth~A Holm, Stephen~M Foiles, and David~L Olmsted.
\newblock {Trends in grain boundary mobility: Survey of motion mechanisms}.
\newblock {\em JOM}, 66(1):114--120, January 2014.

\bibitem{Homer:2015ie}
Eric~R Homer, Srikanth Patala, and J~L Priedeman.
\newblock {Grain Boundary Plane Orientation Fundamental Zones and
  Structure-Property Relationships }.
\newblock {\em Sci. Rep.}, 5:15476, 2015.

\bibitem{Kiyohara:2015wb}
Shin Kiyohara, Tomohiro Miyata, and Teruyasu Mizoguchi.
\newblock {Prediction of grain boundary structure and energy by machine
  learning}.
\newblock December 2015.

\bibitem{lemaitre2016imbalanced}
Guillaume Lema\^{i}tre, Fernando Nogueira, and Christos~K. Aridas.
\newblock Imbalanced-learn: A python toolbox to tackle the curse of imbalanced
  datasets in machine learning.
\newblock {\em CoRR}, abs/1609.06570, 2016.

\bibitem{Lu:2004gh}
L~Lu.
\newblock {Ultrahigh Strength and High Electrical Conductivity in Copper}.
\newblock {\em Science}, 304(5669):422--426, April 2004.

\bibitem{Meyers:2006co}
M~A Meyers, A~Mishra, and D~J Benson.
\newblock {Mechanical properties of nanocrystalline materials}.
\newblock {\em Progr Mat Sci}, 51(4):427--556, May 2006.

\bibitem{Nguyen:2011:BOI:1972030.1972031}
Hien~M. Nguyen, Eric~W. Cooper, and Katsuari Kamei.
\newblock Borderline over\&\#45;sampling for imbalanced data classification.
\newblock {\em Int. J. Knowl. Eng. Soft Data Paradigm.}, 3(1):4--21, April
  2011.

\bibitem{Olmsted:2009ge}
David~L Olmsted, Stephen~M Foiles, and Elizabeth~A Holm.
\newblock {Survey of computed grain boundary properties in face-centered cubic
  metals: I. Grain boundary energy}.
\newblock {\em Acta Materialia}, 57(13):3694--3703, August 2009.

\bibitem{Olmsted:2009in}
David~L Olmsted, Elizabeth~A Holm, and Stephen~M Foiles.
\newblock {Survey of computed grain boundary properties in face-centered cubic
  metals-II: Grain boundary mobility}.
\newblock {\em Acta Materialia}, 57(13):3704--3713, August 2009.

\bibitem{Petch:1953ws}
N~J Petch.
\newblock {The Cleavage Strength of Polycrystals}.
\newblock {\em J Iron Steel Inst}, 174(1):25--28, 1953.

\bibitem{Read:1950um}
WT~Read and W~Shockley.
\newblock {Dislocation Models of Crystal Grain Boundaries}.
\newblock {\em Phys Rev}, 78(3):275--289, 1950.

\bibitem{Shimada:2002jn}
M~Shimada, H~Kokawa, Z~J Wang, Y~S Sato, and I~Karibe.
\newblock {Optimization of grain boundary character distribution for
  intergranular corrosion resistant 304 stainless steel by twin-induced grain
  boundary engineering}.
\newblock {\em Acta Materialia}, 50(9):2331--2341, May 2002.

\bibitem{Spearot:2008bq}
Douglas~E Spearot.
\newblock {Evolution of the E structural unit during uniaxial and constrained
  tensile deformation}.
\newblock {\em Acta Materialia}, 35(1-2):81--88, January 2008.

\bibitem{Sutton:1989vz}
A~P Sutton.
\newblock {On the structural unit model of grain boundary structure}.
\newblock {\em Phil Mag Lett}, 59(2):53--59, 1989.

\bibitem{Sutton:1995ux}
AP~Sutton and RW~Balluffi.
\newblock {\em {Interfaces in Crystalline Materials}}.
\newblock Oxford University Press, Oxford, 1995.

\bibitem{Tadmor:2011tb}
Ellad~B Tadmor and Ronald~E Miller.
\newblock {\em {Modeling Materials}}.
\newblock Continuum, Atomistic and Multiscale Techniques. Cambridge University
  Press, Cambridge, 2011.

\bibitem{Tschopp:2007wn}
M~A Tschopp, G~J Tucker, and D~L McDowell.
\newblock {Structure and free volume of symmetric tilt grain boundaries with
  the E structural unit}.
\newblock {\em Acta Materialia}, 2007.

\bibitem{Tschopp:2007hr}
Mark~A Tschopp and David~L McDowell.
\newblock {Structural unit and faceting description of Sigma 3 asymmetric tilt
  grain boundaries}.
\newblock {\em J Mater Sci}, 42(18):7806--7811, September 2007.

\bibitem{Weins:1969dp}
M~Weins, B~Chalmers, H~Gleiter, and M~ASHBY.
\newblock {Structure of high angle grain boundaries}.
\newblock {\em Scripta Metall Mater}, 3(8):601--603, August 1969.

\bibitem{Wolf:1990ud}
Dieter Wolf.
\newblock {A broken-bond model for grain boundaries in face-centered cubic
  metals}.
\newblock {\em J Appl Phys}, 68(7):3221--3236, 1990.

\bibitem{Wolf:1990fm}
Dieter Wolf.
\newblock {Correlation between structure, energy, and ideal cleavage fracture
  for symmetrical grain boundaries in fcc metals}.
\newblock {\em J Mater Res}, 5(08):1708--1730, August 1990.

\bibitem{Wolf:1990fk}
Dieter Wolf.
\newblock {Structure-Energy Correlation for Grain Boundaries in FCC
  metals{\textemdash}III. Symmetrical Tilt Boundaries}.
\newblock {\em Acta metallurgica et Materialia}, 38(5):781--790, May 1990.

\bibitem{Wolf:1992ve}
Dieter Wolf and Sidney Yip, editors.
\newblock {\em {Materials Interfaces: Atomic-level structure and properties}}.
\newblock Chapman {\&} Hall, London, 1992.

\bibitem{Yang:2010bs}
J~B Yang, Y~Nagai, and M~Hasegawa.
\newblock {Use of the Frank{\textendash}Bilby equation for calculating misfit
  dislocation arrays in interfaces}.
\newblock {\em Scripta Mater}, 62(7):458--461, April 2010.

\end{thebibliography}
\end{document}